\documentclass[usenatbib]{mn2e}
\usepackage{natbibmnfix,graphicx,times}

\newcommand{\bq}{\begin{equation}}
\newcommand{\eq}{\end{equation}}
\newcommand{\bqa}{\begin{eqnarray}}
\newcommand{\eqa}{\end{eqnarray}}

\newcommand{\cmden}{\mbox{ cm$^{-3}$}}
\newcommand{\Mpc}{\mbox{ Mpc}}

\newcommand{\keV}{\mbox{ keV}}
\newcommand{\eV}{\mbox{ eV}}
\newcommand{\kel}{\mbox{ K}}

\newcommand{\yr}{\mbox{ yr}}

\newcommand{\lya}{Ly$\alpha$ }
\newcommand{\lyans}{Ly$\alpha$} 

\title[Secondary ionization and heating by fast electrons]{Secondary ionization and heating by fast electrons}

\author[S.~R. Furlanetto \& S.~Johnson Stoever]{Steven R.  Furlanetto$^{1,2}$\thanks{Email:  sfurlane@astro.ucla.edu} \& Samuel Johnson Stoever$^{1,3}$\\
$^1$Department of Physics \& Astronomy, University of California Los Angeles; Los Angeles, CA 90095, USA\\
$^2$NASA Lunar Science Institute, NASA Ames Research Center, Moffett Field, CA\\
$^3$Department of Astronomy, Cornell University, 610 Space Sciences Building, Cornell University, Ithaca, NY 14853}

\begin{document}

\maketitle

\begin{abstract}
We examine the fate of fast electrons (with energies $E>10 \eV$) in a thermal gas of primordial composition.  To follow their interactions with the background gas, we construct a Monte Carlo model that includes: (1) electron-electron scattering (which transforms the electron kinetic energy into heat), (2) collisional ionization of hydrogen and helium (which produces secondary electrons that themselves scatter through the medium), and (3) collisional excitation (which produces secondary photons, whose fates we also follow approximately).  For the last process, we explicitly include all transitions to upper levels $n \le 4$, together with a well-motivated extrapolation to higher levels.  In all cases, we use recent calculated cross-sections at $E<1 \keV$ and the Bethe approximation to extrapolate to higher energies.  We compute the fractions of energy deposited as heat, ionization (tracking HI and the helium species separately), and excitation (tracking HI \lya separately) under a broad range of conditions appropriate to the intergalactic medium.  The energy deposition fractions depend on both the background ionized fraction and the electron energy but are nearly independent of the background density.  We find good agreement with some, but not all, previous calculations at high energies.  Electronic tables of our results are available on request.
\end{abstract}
  

\section{Introduction} \label{intro}

Fast electrons, typically generated by high-energy photons or cosmic ray collisions, are crucial for a wide range of astrophysical problems.  For example, cosmic ray ionization is important for the thermal balance of interstellar clouds \citep{spitzer69}, electrons produced by the Comptonization of $\gamma$-ray photons affect the early history of supernova remnants \citep{xu91b}, and broad-line emission regions around quasars are likely ionized by hard photons from those sources, making the fate of the liberated electrons of paramount importance \citep{shull85}.  

As such, there have been numerous studies of the interactions between these electrons and a background gas, through the processes of collisional excitation, ionization, and electron-electron scattering.  These calculations have included both analytic approaches \citep{spitzer68, spitzer69, jura71, bergeron73, xu91} and numerical explorations \citep{habing71, shull79, shull85, valdes08}.  Most have focused on interactions with atomic or ionic gases, relevant especially to low-density material in the interstellar (or intergalactic) media.  Others have considered the additional effects of molecules (e.g., \citealt{dalgarno99}).

Recently, the fate of X-rays in the high-redshift intergalactic medium (IGM) has become an important question.  Before the reionization of HI, ultraviolet ionizing photons are trapped near their sources, but X-rays can travel much larger distances through the IGM.  As such, they are thought to provide the most important radiation background for the bulk of the IGM, slowly ionizing the mostly  neutral gas far from stars.  However, most of their energy is deposited indirectly, through the fast electrons generated by photo-ionization.  For example, the electrons produced by X-rays from the first hard sources (quasars, supernova remnants, or even Population III stars; \citealt{oh01, venkatesan01}) are most likely responsible for heating the IGM to $T \sim 1000 \kel$ before reionization  \citep{kuhlen05-sim, furl06-glob}.  

This heating and ionization could influence future generations of structure (e.g., \citealt{ricotti02, oh03-entropy}) and it has important observational consequences.  For example, it can dramatically change the highly-redshifted 21 cm signal from the early IGM \citep{furl06-review, kuhlen06-21cm, pritchard07}, and it could even provide a signature of exotic physics during the Dark Ages before the first sources appear \citep{chen04-decay, furl06-dm, mapelli06, valdes08}.  Moreover, HI \lya photons produced as the secondary electrons collisionally excite atoms in the background gas can greatly affect the 21 cm signal by changing the hyperfine level populations of HI \citep{chuzhoy06-first, pritchard07, chen08}, and the secondary ionizations may play an important role in the HI (and HeI) reionization of the IGM \citep{ricotti04_a, ricotti05, volonteri09}.

Here we revisit this problem using a Monte Carlo model (described in \S \ref{mc}) that incorporates the most recent cross-sections for interactions between X-rays and primordial gas.  We review the relevant physics of collisional ionization, excitation, and electron scattering in \S \ref{ion}-\ref{heat} and comment on some of our approximations in \S \ref{ignored}.  We present our principal results, together with a comparison to previous work, in \S \ref{results} and a simple example of their utility for heating of the high-redshift IGM in \S \ref{example}.  Finally, we conclude in \S \ref{disc}.

\section{The Monte Carlo model}
\label{mc}

Because of the wide range of interactions available to fast electrons, we use a Monte Carlo model to track their fates.  Our procedure is very similar to \citet{shull85} (see also \citealt{shull79} and \citealt{valdes08}), except that we use updated cross-sections and include (and track) more interaction processes.  An alternative, fully analytic, approach uses the degradation equation \citep{spencer54, xu91}, but we find that to be more cumbersome.

As inputs, the model requires only the number densities and ionized fractions of hydrogen and helium.  We will see that the ionized fractions have an important effect on the results but that the absolute number densities have only a slight impact (through the Coulomb logarithm).  For simplicity, we will typically assume that the HI and HeI fractions are identical ($x_i$) and that the HeII fraction is $1-x_i$.  The presence of HeIII does not significantly affect our results.  As a fiducial value, we use the mean density of the Universe at $z=10$ (according to the cosmological parameters recommended by \citealt{dunkley09}), but taking any density $n \la 10^8 \cmden$ only changes our reported values by a few percent.  Note that our model assumes a static IGM and does not self-consistently account for the ionization (and heating) produced by each electron.

The model begins with a single electron of energy $E >10.2 \eV$; lower energy particles can only interact with the electron gas and so automatically deposit all of their energy as heat.  Our goal is to compute the fraction of this energy that is deposited in ionizing each of the relevant IGM species (HI, HeI, or HeII), the fraction of energy that is lost to collisional excitations producing photons with $E<13.6 \eV$ (and additionally the fraction that ends up in HI \lya photons), and the fraction of energy that is deposited as heat in the IGM.  We also follow collisional excitations that produce higher energy photons (from helium), but the resulting photons are re-absorbed by the IGM and so are not the ultimate products of the process.

To this end, we compute the cross-sections (weighted by the number densities of each species) for the electron to interact with the IGM in several different ways:  collisional ionization (of any of the three species above), collisional excitation (again of any of the three species, and including explicitly all levels $n \le 4$), and electron-electron collisions.  Below we describe our treatment of each of these processes in detail.  From these cross-sections we randomly choose one such process for the electron to undergo and update its energy, and the energy deposition fractions, accordingly.  If it ionizes an atom or ion, we add a new secondary electron of the appropriate energy to the array of particles; if it collisionally excites helium, we add a new photon to the array.\footnote{Collisional excitation of hydrogen produces a photon with $E<13.6 \eV$ that no longer affects the IGM.}  We repeat this process until the primary electron's energy falls below 10.2 eV; beyond that point, the only available process is a collision with an electron, so we assign all its energy to heat.

We then track each of the secondary electrons through the same machinery, and finally we randomly determine (from the photoionization cross-sections weighted by number density) the species that each secondary photon ionizes and follow the resulting electrons through their entire energy loss cascade.  At each step, we track the energy lost to each of the aforementioned processes.

For our final results, we follow $10^5$ input electrons at each of 258 energies (logarithmically spaced from 10 eV to 9900 eV) at each of fourteen ionized fractions from $x_i=10^{-4}$ to 0.999.  This encompasses the range expected in the early IGM (where $x_i$ is set by the relic density following recombination; \citealt{seager99}) up to the point at which the neutrals are no longer significant energy sinks.  

To generate random numbers, we use the Mersenne Twister algorithm, which has a period of at least $2^{19937}-1$ \citep{matsumoto98}.

\section{Collisional ionization}
\label{ion}

For $E<1 \keV$, we take the collisional ionization cross-sections from the CCC database,\footnote{See http://atom.curtin.edu.au/CCC-WWW/.} an online collection of cross-sections calculated with the convergent close-coupling (CCC) method.  The CCC approach is accurate whenever the target particle can be well-modelled by one or two valence electrons above a Hartree-Fock core; obviously this is an excellent approximation for hydrogen and helium.  The relevant physics, and references to many of the original papers for hydrogen and helium, can be found in \citet{bray02}.  We use cubic splines to interpolate the database values.

At higher energies, we assume that cross-sections follow the Bethe approximation limit \citep{bethe30}.  Then the cross-section for ionization from the ground state in species $i$ is
\bq
\sigma_{i} \sim {2 E_i \over E} \left( A_i \ln {E \over 2 E_i} + B_i \right) \pi a_0^2,
\eq
where $a_0$ is the Bohr radius and $A_i$ and $B_i$ are coefficients.  These are usually fixed by demanding that the cross-section map onto the first Born approximation at high energies.  This asymptotic behavior compares well to the calculated behavior at $E \sim 1 \keV$, and we extrapolate to higher energies by fitting a function of this form to the uppermost energy bin in the CCC results.  The resulting parameters $A_H$ and $B_H$ differ from the analytic estimates of \citet{johnson72} by $\sim 15\%$ and $30\%$, respectively, but their exact values make little difference to our final results (largely because it is the relative importance of each ionization and excitation process that matters).

Whenever a species is collisionally ionized, it also produces a secondary electron, so we must select the final energies of the incident and ejected electrons.  We use the probability distributions of \citet{dalgarno99}, which are adapted from the measurements of \citet{opal71}; see also \citet{shull79} for a discussion of the energy spectra of secondaries.  This prescription assumes that the probability for the secondary to have an energy $\varepsilon$ is proportional to
\bq
p(\varepsilon) \propto {1 \over 1+ (\varepsilon/\bar{\varepsilon}_i)^{2.1}},
\label{eq:psec}
\eq
where $\bar{\varepsilon}_i= 8,\,15.8$, and $32.6 \eV$ for HI, HeI, and HeII, respectively.  (We always identify the ``secondary" as having the lower energy of the two final photons, so that $\varepsilon < (E-E_i)/2$.)  Note that secondary electrons are typically ejected with a modest energy somewhat below the ionization threshold of their original host atom:  the median energies are $7.2,\,14.2$, and $28.5 \eV$ for the three species, although there is a tail to much higher $\varepsilon$.  As discussed in \citet{shull79}, the \emph{mean} secondary energy increases logarithmically with the incident energy for fast collisions.\footnote{Note that the exponent of 2.1 and coefficients $\varepsilon_i$ in equation~(\ref{eq:psec}) differ slightly from the fits of \citet{shull79}.}

\citet{johnson72} provides an alternate estimate of the secondary electron energy distribution for HI by constructing analytic approximations to the HI ionization and excitation cross-sections.  In his calculation, the leading order behavior is $\propto (1+E/E_i)^{-2}$, although he also includes several other corrective terms proportional to higher powers of this same factor (see also \citealt{omidvar69}).  This has a similar (though not identical) shape to our choice, and there is no simple way to extend it to the other relevant ions.

\section{Collisional excitation}
\label{excite}

We include collisional excitation processes in a similar way to collisional ionization, by making use of the CCC database.  This provides cross-sections for all excitations from the $n=1$ to $n=2,\,3,$ and 4 states of HI, HeI, and HeII, including separately all angular momentum sublevels as well as separate estimates for the singlet and triplet configurations in the case of HeI.  We again extrapolate to $E>1 \keV$ using the Bethe approximation.

This explicit separation is useful to us because some applications (particularly estimates of the high-redshift 21 cm signal) require the production rate of HI \lya photons through collisional excitation.  For the $n < 4$ levels, this is relatively easy (see \citealt{pritchard06} and \citealt{hirata06} for discussions of analogous photo-excitation processes).\footnote{Note that the discussion below ignores the possibility of collisional de-excitation and so will break down in sufficiently dense gas.}  Atoms excited to the $2s$ state obviously produce no \lya photons.  Atoms excited to the $3s$ or $3d$ levels must either decay through the $2p$ state (producing a \lya photon) or directly to the ground state.  But in the latter case the resulting line photon will quickly be re-absorbed by a nearby atom (assuming that neutrals are encountered before the photon redshifts out of resonance), and it will eventually cascade into a \lya photon.  On the other hand, atoms excited to the $3p$ state must decay to $2s$ and then to the ground state via two-photon decay.  Thus they do not produce \lya photons at all.

The cascade possibilities become more complicated at $n=4$.  For example, excitation to $4s$ can result in decay directly to the ground state (which is irrelevant for our purposes, since the photon is re-absorbed) or indirectly through the $3p$ state (in which case no \lya photon is produced) or the $2p$ state (which produces a \lya photon).  The relative fractions of these are given by the decay probabilities, 
\bq
p_{\rm Ly\alpha}(4s) = { A^e_{4s-2p} \over A^e_{4s-2p} + A^e_{4s-3p}},
\label{eq:plya}
\eq
where $A^e_{nm}$ is the spontaneous emission coefficient from level $n$ to $m$.  Similar exercises yield the decay probabilities for the other $n=4$ sublevels.  It is important to include excitations to the different $n \ge 2$ angular momentum sublevels separately, because they feed into \lya differently; most previous work has included only $n \le 2$ (e.g., \citealt{shull85}), averaged over all the sublevels (e.g., \citealt{xu91}), or included only excitation to the $np$ sublevels (e.g., \citealt{valdes08}).

Note that, in the case of HeI and HeII excitations, the decay photons can ionize other elements. In detail, we should follow the radiative cascades of these excited atoms or ions, propagate each decay photon through the IGM, and follow the resulting ionizations.  However, we take a cruder approach and simply assume that each excited atom or ion immediately decays to the ground state, producing a photon equal to the energy of the excitation.  We then assume that this photon ionizes another atom and follow the resulting photo-electron.  We use the photoionization cross-sections from \citet{verner96} for this purpose.  

We therefore eventually recycle \emph{all} the HeI and HeII excitation energy into ionization or heating, rather than allowing a fraction to escape.  However, we find that $\la 1.5\%$ of the initial electron energy is typically lost to HeI excitation (and even smaller than that for HeII) in the high energy limit, and most of that is direct excitation to $2p$, for which our assumption is accurate.  This approximation is therefore a small correction to our results (comparable to many other processes that we ignore, and smaller than uncertainties in our cross-sections).  It will, however, affect the results more strongly near the HeI threshold (e.g., Fig.~1 of \citealt{shull85}).

\subsection{Excitation to $n>4$ states}
\label{highexcite}

The CCC database does not include cross-sections for transitions to $n>4$ states, and in any case it is impractical to include arbitrarily many final states.  We therefore approximate the effects of higher levels based on the analytic calculations of \citet{johnson72} and \cite{xu91}.

In the Bethe approximation, the leading coefficient $A_{n}$ (for excitation from the ground state to the $n$th level) is proportional to the oscillator strength of the equivalent radiative absorption transition, $f_n$.  This depends on the quantum number of the upper level through $f_n \propto g_n/(n k_n)^{3}$, where $k_n=1-(1/n)^2$ and $g_n$ is a correction factor (the ``Gaunt factor") that depends on the quantum mechanical details of the interaction.  Thus we can approximate the total cross-section to all levels above $n=4$, relative to that for $n=4$, as
\bq
{\sigma_{n \ge 5} \over \sigma_{n=4}} = 60 \sum_{n=5}^\infty (k_n n)^{-3} \approx 1.58.
\label{eq:ngt4}
\eq

Because of the details that we have ignored (namely the Gaunt factor), the actual cross-section of the higher levels is somewhat smaller than this estimate.  \citet{xu91} explicitly included all $n \le 10$ in their calculation (utilizing the analytic approximate cross-section of \citealt{johnson72}); they found that the $n=5$--10 states increased the interaction rate by a factor of 1.17, as opposed to the 1.29 suggested by truncating the sum in equation~(\ref{eq:ngt4}) at $n=10$.  We therefore increase the $n=4$ cross-sections by a factor $(1.17/1.29) \times 1.58 = 1.43$ for a fiducial estimate of the importance of higher-level transitions.  Fortunately, varying this enhancement level by a factor of two affects our heating, ionization, and excitation fractions by less than a percent. 

We further assume that the fraction of such excitations producing \lya photons is identical to that for $n=4$, which is a reasonable estimate \citep{pritchard06} but does not substantially affect our results anyway.

We also apply this same $n>4$ enhancement to HeI and HeII, although this is no more than a guess.  Again, this makes only a negligible difference to our results.

\section{Electron-electron collisions and heating}
\label{heat}

A fast electron will share its kinetic energy with its surroundings by scattering off of ambient electrons.  The energy loss rate is \citep{spitzer69}
\bq
{dE \over dt} = - {4 \pi e^4 n_e \ln \Lambda \over m_e w}, 
\label{eq:energyloss}
\eq
where $\ln \Lambda$ is the Coulomb logarithm (to be discussed in \S \ref{coul-log} below) and $w$ is the velocity of the electron.  However, this near-continuous slowing-down (thanks to the long-range interactions of the Coulomb force) must be discretized to be included in our Monte Carlo formalism.

We follow \citet{habing71} and \citet{shull79} by casting this process in the following form.  First, we assume that the electron loses a fixed fraction $f$ of its energy in each interaction.  We then rewrite the energy loss rate as
\bq
{dE \over dt}  \approx {f E \over t_{\rm coll}} = fE \times w n_e \sigma_{ee},
\label{eq:energyloss2}
\eq
where the collision time is $t_{\rm coll} = 1/(w n_e \sigma_{ee})$ and we have let the interaction cross-section be $\sigma_{ee}$.  Rearranging, and taking the non-relativistic limit $E=mw^2/2$, we obtain
\bq
\sigma_{ee} = {40 \pi e^4 \over E^2} \ln \Lambda \left( { 0.05 \over f} \right).
\label{eq:eecs}
\eq

Although the parameter $f$ is artificial, it does not affect our calculations:  all models with $f<0.05$ yield identical results, within the errors expected in our Monte Carlo simulations.

\section{Approximations}
\label{ignored}

Before proceeding to our results, we now discuss a few subtleties that arise in the calculation.  In addition to the processes that we discuss in detail below, our model also ignores several subdominant mechanisms that have been studied before; we list them here for completeness.  First, we ignore the double ionization of neutral helium, which occurs at a rate $\sim 2\%$ that of single ionization \citep{dalgarno99}.  We also ignore Coulomb collisions with protons and recombinations.  \citet{valdes08} included these two latter processes and showed that they are negligible.

\subsection{Timescales}
\label{timescale}

Our model assumes that all of the photon's energy is deposited in the IGM instantaneously.  In situations where that energy deposition occurs over a long time period, this may not be a good approximation (for example, if the density field or ionized fraction evolve rapidly compared to the energy deposition timescale).  We can estimate the relevant timescales by considering electron-electron interactions and collisional ionization of HI, which are typically the most important processes.  For the former, we find that 
\bq
t_{\rm loss,e} = \left| {E \over dE/dt} \right| \sim 5 \times 10^3 x_i^{-1} \left( {E \over {\rm keV}} \right)^{3/2} \left( {1+z \over 10} \right)^{-3} \yr,
\label{eq:tloss-e}
\eq
where we have used the Coulomb logarithm appropriate to the IGM at $z=10$.  Clearly, even with an ionized fraction $x_i \ga 10^{-3}$, this timescale is only a small fraction of the Hubble time, so the primary photoelectron will rapidly lose its energy once it is created.  

At smaller ionized fractions, electron scattering may be slow.  However, with the Bethe approximation form for the collisional ionization cross-section, and assuming that $\Delta E \sim E_H$ in each interaction, the energy loss timescale from ionization is
\bq
t_{\rm loss,H} \sim 5 \times 10^5 x_H^{-1}  \left( {E \over {\rm keV}} \right)^{3/2} \left( {1+z \over 10} \right)^{-3} \yr,
\label{eq:tloss-h}
\eq
where $x_H = 1 - x_i$ and we have evaluated the logarithmic factor at $E=1 \keV$ as well.  The two processes are comparable when $x_i \sim 0.01$.  Thus $t_{\rm loss} \la 5 \times 10^5 \yr$ for keV photons:  much smaller than the Hubble time, so our instantaneous approximation is reasonable.  The only exceptions are very high energy electrons (which result only from high energy X-rays, to which the Universe is transparent anyway).

Of course, if the underlying ionization fraction changes rapidly even this timescale may not be short enough.  While such changes are unlikely to occur faster than the Hubble time on a global scale, the ionized fraction can certainly change rapidly on a local level when a bright source ionizes its environs.  We therefore urge caution in using our results under such circumstances.

\subsection{Inverse Compton scattering}
\label{compton}

We have also ignored energy losses due to inverse Compton scattering from CMB photons.  The emitted power due to this process is $dE/dt = - (4/3) \sigma_T c U_{\rm rad} \gamma^2 \beta^2$, where $\sigma_T$ is the Thomson cross-section, $U_{\rm rad} = a T_{\rm CMB}^4$ is the energy density of the CMB, $\beta = w/c$, and $\gamma^2 = 1/(1-\beta^2)$ \citep{rybicki79}.  For a non-relativistic electron, the timescale for inverse Compton cooling is therefore
\bq
t_{\rm comp} = {3 m_e c \over 8 \sigma_T a T_{\rm CMB}^4} \approx 10^8 \left( {1 + z \over 10} \right)^{-4} \yr.
\label{eq:tcomp}
\eq
Note that this is independent of energy, so inverse Compton cooling is most important for high-energy electrons, where the other loss timescales are large.

Because $t_{\rm comp}$ is typically much longer than the loss timescale due to collisional processes (which is bounded from above by eqs.~\ref{eq:tloss-e} and~\ref{eq:tloss-h}), and because of the strong redshift dependence, we have ignored inverse Compton scattering in our calculations.  It can be roughly included by assuming that a fraction $t_{\rm loss}/t_{\rm comp}$ of the electron energy is lost to the CMB and reducing the other fractions accordingly (note that their proportions will not change significantly, unless $x_i \ll 10^{-2}$ and the energy is near a line threshold).

\subsection{The IGM density and temperature}
\label{coul-log}

In all of our calculations, we assume that the background gas has a density equal to the cosmic mean at $z=10$.  But, to a good approximation, our results are independent of the absolute density, because the interaction rates all scale linearly with it.  However, the Coulomb logarithm in equation~(\ref{eq:eecs}) does have an implicit density dependence that breaks this convenient feature.  If one considers only the scattering of the photoelectron from discrete particles, the appropriate Coulomb logarithm is \citep{spitzer62}
\bq
\ln \Lambda = \sqrt{ (k T)^3 \over \pi n_e e^6 \gamma_e},
\label{eq:coul-log-naive}
\eq
where $n_e$ is the electron density, $T$ is the gas temperature, and $\ln \gamma_e$ is Euler's constant.  In this regime, this factor is independent of the electron's energy, so $\sigma_{ee} \propto E^{-2}$ and heating rapidly becomes less important at high energies compared to ionization and excitation.

However, the photoelectron can also have collective interactions with the plasma as a whole.  This process becomes more important at high energies. The effective Coulomb logarithm, including both individual collisions and  these collective interactions, is \citep{schunk71}
\bq
\ln \Lambda = \ln \left[ {2 E \over \hbar} \left( {4 \pi n_e e^2 \over m_e} \right)^{-1/2} \right] \equiv  \ln \left( {4 E \over \zeta_e} \right).
\label{eq:coul-log}
\eq
so long as $E > me^4/2 \hbar^2 \approx 13.7 \eV$.\footnote{We are interested in slightly lower energies as well, but the dynamics there are almost completely dominated by the structure of the line spectrum of HI rather than the details of $\sigma_{ee}$.}  Here, the factor $\zeta_e = 7.40 \times 10^{-11} (n_e/{\rm cm}^{-3})$ eV.  In this case, $\sigma_{ee} \propto E^{-2} \ln E$, so heating is somewhat more important at high energies, although $\sigma_{ee}$ still falls off by one power of energy faster than the ionization and excitation cross-sections.  Note that, when plasma effects are included, the cross-section is independent of the temperature of the gas, so we need not specify it in our calculations. 

We use equation~(\ref{eq:coul-log}) in our fiducial calculations, following \citet{xu91}.  Ignoring collective effects reduces the fraction of incident energy deposited as heat at $E \ga 100 \eV$ by a few percent, and it correspondingly increases the fractions of energy deposited in ionization and excitation. 

The density can also play a second role in collisional de-excitation of the target atoms and in populating the upper levels of the atoms (provided also that the temperature is sufficiently large).  This requires $n \gg 10^8 \cmden$ so is not important for the IGM.  But we caution readers against trusting our results in such environments.

\section{Results}
\label{results}

We now present our main results.  As a reminder, we use $10^5$ Monte Carlo trials for each electron energy (which ranges from $10$--9900~eV), and we explicitly follow all secondary electrons and photons.  We take $f=0.05$ and set the density of the background gas to be the mean cosmic density at $z=10$, but we saw above that the results are very insensitive to these parameters.  The primary input parameters are then the initial electron energy and the ionization fractions.  For the latter, we let $x_i$ be the density of HII relative to the total hydrogen density and assume that this is also the fraction of helium in the form of HeII, with zero HeIII.

We present our results in terms of energy deposition fractions, with $f_{\rm ion}$, $f_{\rm heat}$, and $f_{\rm excite}$ the fractions of the initial \emph{electron} energy that goes into ionization, heating, and HI line photons generated by collisional excitation.\footnote{We caution the reader that our parameters do not include the effects of the initial ionization event that generates the electron.}  (Note that line photons from helium are transformed into ionizing photons and so deposit their energy in other ways, implicit in our code.)  We also let $f_{\rm Ly\alpha}$ be the fraction of energy deposited in HI \lya photons.  We have verified explicitly that our code conserves energy throughout the entire interaction cycle.

\begin{figure}
\begin{center}
\resizebox{8cm}{!}{\includegraphics{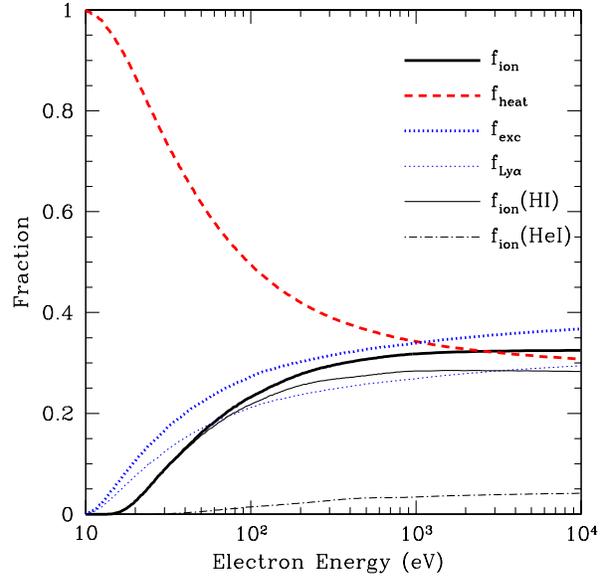}}\\%
\end{center}
\caption{Energy deposition fractions as a function of initial electron energy, for an IGM with $x_i=0.01$.  The thick solid curve shows the fraction of energy used in ionization; the thin solid and dot-dashed curves show the fractions used for ionizing HI and HeI, respectively.  The thick dashed curve shows the fraction deposited as heat through electron-electron interactions.  The thick dotted curve shows the total energy lost to HI line photons; the thin dotted curve shows the fraction in HI \lya photons.}
\label{fig:xh099}
\end{figure}

Figure~\ref{fig:xh099} shows these fractions, as a function of photon energy, in a medium with $x_i=0.01$.  The thick solid, dashed, and dotted curves show $f_{\rm ion}$, $f_{\rm heat}$, and $f_{\rm excite}$, respectively.  The thin solid and dot-dashed curves show explicitly the fractions of energy going into HI and HeI ionization, respectively.  (The fraction going into HeII ionization is $\ll 1\%$ throughout.)  Finally, the thin dotted curve shows $f_{\rm Ly\alpha}$.

The structure of the curves is relatively easy to understand.  Electrons with $E<10.2 \eV$ are unable to interact with any atoms or ions, so all of the energy is deposited as heat.  As $E$ increases, more and more excitation and ionization processes become available, so the energy injected as heat decreases.  Interestingly, even with this large of a neutral fraction, the individual line thresholds do not introduce discrete features into the deposition fractions, and all of the parameters depend smoothly on energy.

At higher energies, where no additional processes become available, the fractions vary only slowly, eventually approaching reasonably constant values at $E \sim 1$--$10 \keV$, where ionization, heating, and excitation split roughly equally.  As pointed out by \citet{shull85}, the naive expectation from the Bethe approximation (with $\sigma_{i,e} \propto E^{-1} \ln E$ for ionization and excitation and $\sigma_{ee} \propto E^{-2} \ln E$ for heating), that less and less heating should occur at high photon energies, is false.  This is because each ionization produces a moderate energy secondary electron (typically $\sim 10 \eV$, but occasionally much larger).  A large fraction of the primary's energy is lost through these intermediaries, who in turn lose most of their energy to heat, making the behavior at high energies much less variable than one would expect.

\begin{figure*}
\begin{center}
\resizebox{8cm}{!}{\includegraphics{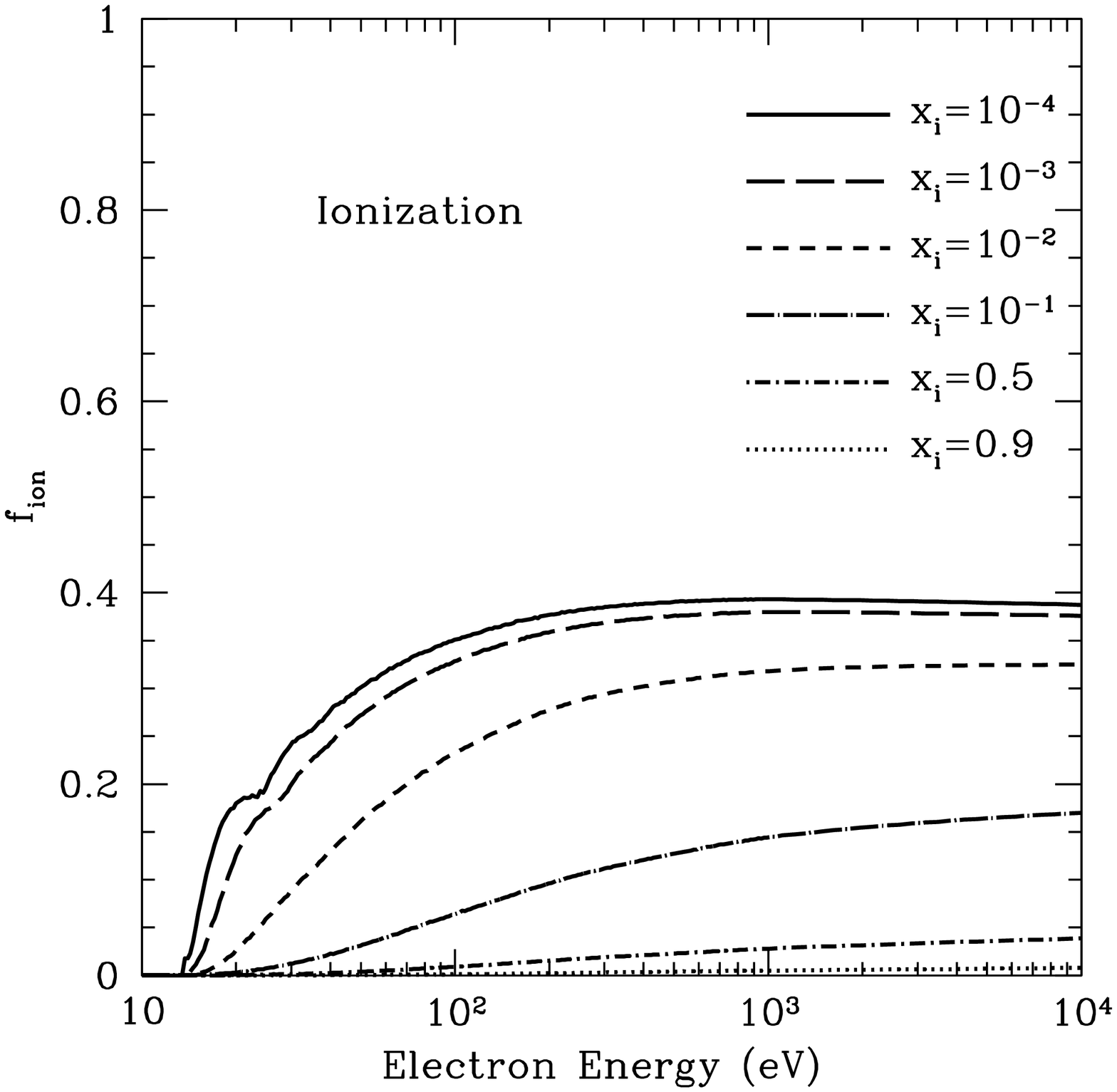}}
\hspace{0.13cm}
\resizebox{8cm}{!}{\includegraphics{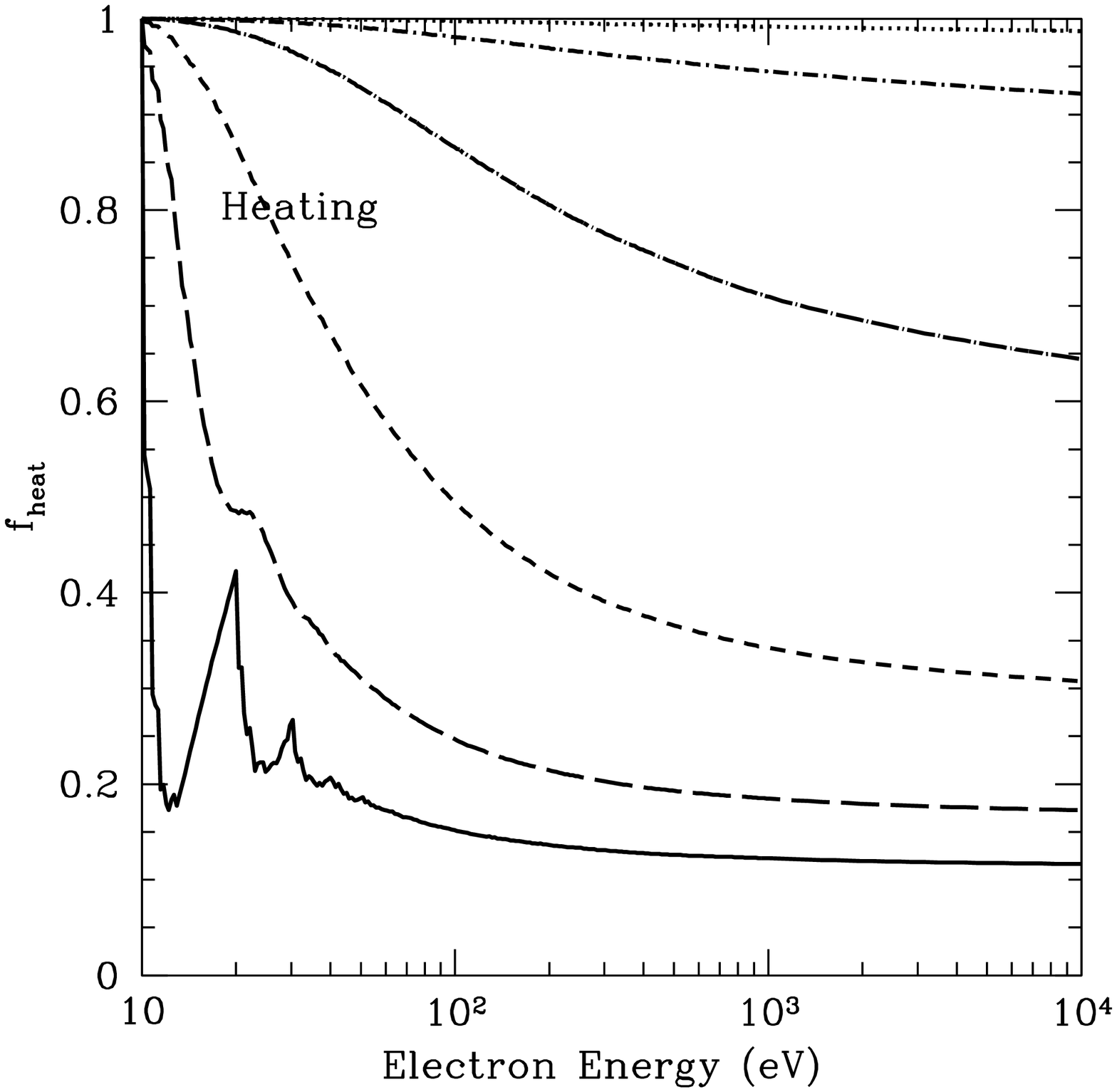}}
\end{center}
\caption{Energy deposition fractions in ionization (\emph{left}) and heating (\emph{right}) as a function of initial electron energy.  The curves show results for several different IGM ionized fractions, as labelled in the left panel.}
\label{fig:ion-heat}
\end{figure*}

\begin{figure*}
\begin{center}
\resizebox{8cm}{!}{\includegraphics{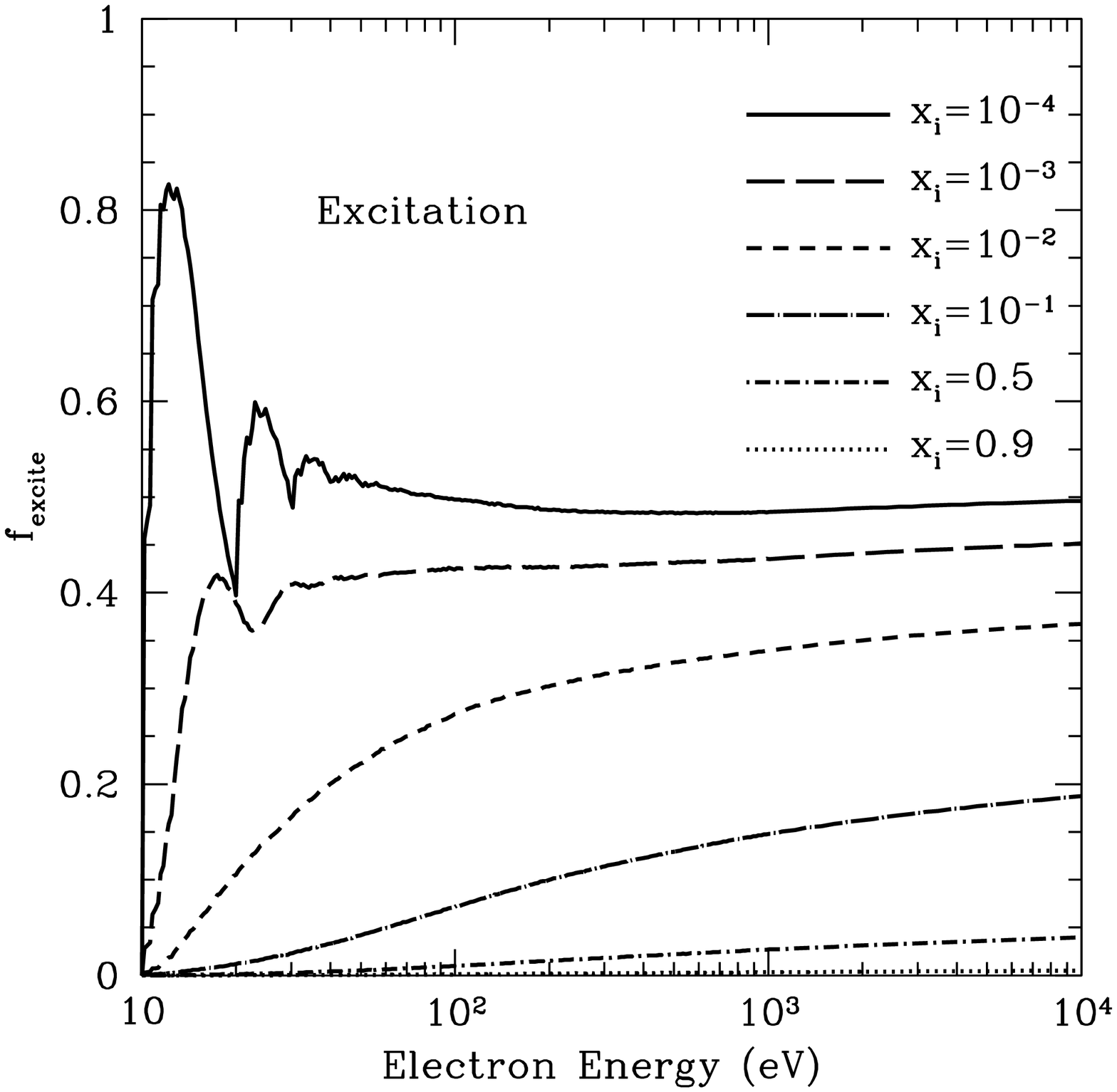}}
\hspace{0.13cm}
\resizebox{8cm}{!}{\includegraphics{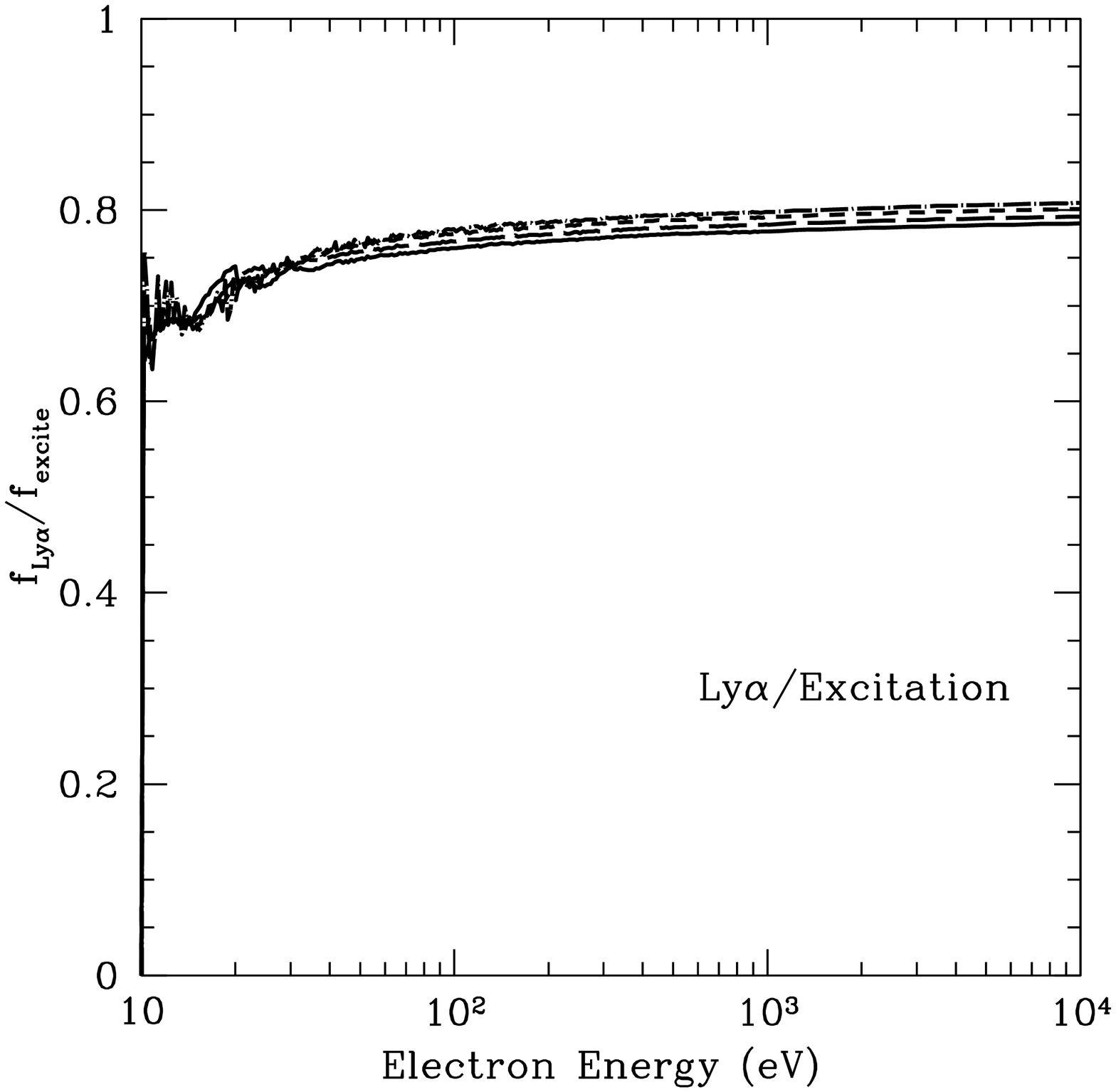}}
\end{center}
\caption{Energy deposition fractions in collisional excitation (\emph{left}) as a function of initial electron energy, and the fraction of this energy that goes into HI \lya photons (\emph{right}).  The curves show results for several different IGM ionized fractions, as labelled in the left panel.  Note that we do not show curves for $x_i=0.5$ or $x_i=0.9$ in the right panel, as they are very noisy at low energies (although the high-energy limit is well-behaved).}
\label{fig:excite-lya}
\end{figure*}

Figures~\ref{fig:ion-heat} and \ref{fig:excite-lya} show the energy deposition fractions for a range of $x_i$.  The first panel shows $f_{\rm ion}$.  For small $x_i$, the ionization thresholds imprint features on the curves.  At high energies, $f_{\rm ion} \sim 0.4$ at small ionized fractions and rapidly approaches zero as $x_i$ increases.  The behavior at very high energies sets an upper limit to the total energy invested in ionization.

The left panel of Figure~\ref{fig:excite-lya} shows the corresponding fractions lost to collisionally excited HI line photons.  Note the line structure apparent at $x_i \le 10^{-3}$; the features are separated by $\sim 10 \eV$ and correspond to multiple excitations of HI atoms.  Although the structure around these features appears noisy, it is real and remains unchanged in other Monte Carlo tests.  However, as noted above these features blend together by $x_i \sim 10^{-2}$.  Note that collisional excitation becomes important at lower photon energies than does ionization, because the lines require less energy; in nearly neutral gas, these lines are clearly the dominant process even near the ionization threshold.  

Interestingly, the right panel of Figure~\ref{fig:excite-lya} shows that the ratio $f_{\rm Ly\alpha}/f_{\rm excite}$ is nearly independent of both energy and $x_i$:  it depends almost entirely on the atomic physics of HI.  The weak energy dependence near threshold occurs because the different levels feed into \lya in different ways.  The (very) weak dependence on $x_i$, with the \lya fraction increasing slightly with $x_i$, is probably because efficient electron-electron interactions in highly-ionized gas bring electrons to the \lya threshold energy more rapidly.  The overall fraction of $\sim 80\%$ can easily be estimated from the oscillator strengths \citep{pritchard07}.

The remaining energy, shown in the right panel of Figure~\ref{fig:ion-heat}, heats the IGM.  The curves vary only slowly at large $E$, but the behavior at $E \la 100 \eV$ is complicated by the ionization and line thresholds.  Note that the features here are caused by the combination of these processes.

\subsection{Comparison to previous work}
\label{comp}

Calculations similar to ours have appeared several times over the past four decades.  The most recent are \citet{shull85}, \citet{xu91}, and \citet{valdes08}.   

\citet{xu91} used the degradation equation \citep{spencer54}, a fully analytic technique, to compute the energy deposition fractions.  They took the approximate (but fully analytic) ionization and excitation cross-sections from \citet{johnson72}, and explicitly included all excitations with $n \le 10$, although only for a pure hydrogen gas.

We have generated a similar set of scenarios to \citet{xu91} for comparison purposes, in which we include only hydrogen target atoms, use their Coulomb logarithm, and extrapolate the excitation cross-sections to only $n \le 10$ levels.  Over the range $x_i=10^{-4}$--$0.9$, we find generally good agreement between the two sets of results at $E=2 \keV$ (where \citealt{xu91} report detailed deposition fractions).  The deviations in the energy deposition fractions are always smaller than a few percent (in absolute terms), with our $f_{\rm heat}$ typically slightly larger and our other fractions slightly smaller; the differences are largest at large ionized fractions.  We note that \citet{dalgarno99} also found slightly larger heating fractions than \citet{xu91}.  At $x_i < 0.01$, all of our results agree to within $1\%$ (again in absolute terms).  Given that we use completely different methods and cross-sections, we regard this agreement as excellent.  There is also good qualitative agreement in the shapes of our energy deposition curves, and especially in the features from excitation and ionization that appear at small $x_i$.

\begin{figure}
\begin{center}
\resizebox{8cm}{!}{\includegraphics{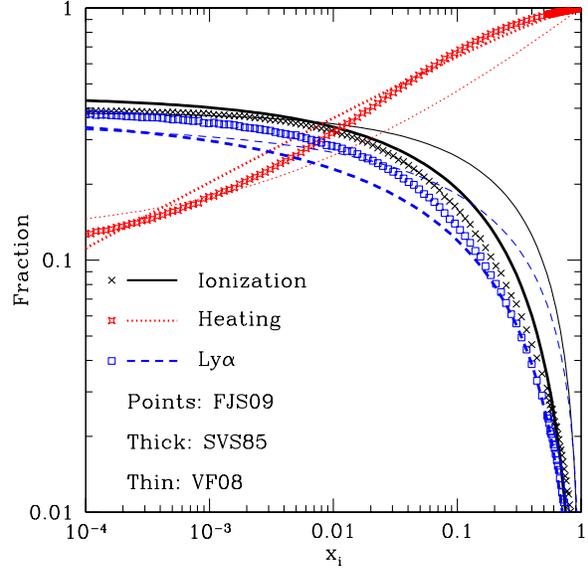}}\\%
\end{center}
\caption{Comparison of our results (points) to previous fitting formulae from \citet{shull85} (thick curves) and \citet{valdes08} (thin curves).  The dotted curves and open diamonds show $f_{\rm heat}$.  The solid and dashed curves (crosses and open squares) show $f_{\rm ion}$ and $f_{\rm Ly\alpha}$, respectively.  Most are evaluated at $E=3 \keV$, except for those from \citet{valdes08}, who use $E=2 \keV$.}
\label{fig:high-energy}
\end{figure}

In contrast, \citet{shull85} used a Monte Carlo method very similar to ours, although we have added several interactions (namely excitation to higher levels) and updated the cross-sections for most of the others (including the Coulomb logarithm in the heating component).  Despite these differences, our results agree to similar (few percent) accuracy in the high-energy limit (or specifically at 3 keV, where \citealt{shull85} provide detailed results).  Figure~\ref{fig:high-energy} provides a detailed comparison as a function of $x_i$; the solid and dashed curves show our results and those from \citet{shull85}, respectively.\footnote{For $f_{\rm Ly\alpha}$ we have used their expression for the fraction of energy deposited in collisional excitation of HI, multiplied by a factor 0.8 as in Figure~\ref{fig:excite-lya}.}

As one might expect given their lack of data for excitations to high $n$ states, \citet{shull85} somewhat underestimate the importance of collisional excitation, which can cause a discrepancy of up to $\sim 6\%$ with respect to our results at small neutral fractions.  They overestimate $f_{\rm ion}$ by a similar amount compared to our calculations, but $f_{\rm heat}$ agrees rather well (although our results clearly have more curvature than theirs).  Overall, we regard this agreement as quite good.

\citet{valdes08} also used a Monte Carlo model to examine the energy deposition fractions for high-energy electrons, with $E>3 \keV$.  They included most of the same processes we have, albeit with different cross-sections in most cases.  The most significant difference is their treatment of collisional excitation; they took values from \citet{stone02}, who computed their cross-sections with the scaled plane-wave Born approximation.  \citet{stone02} show explicit comparisons with the CCC database; the CCC values differ by $\la 20\%$.  However, most significantly \citet{stone02} only compute excitation cross-sections to the $np$ angular momentum sublevels.  Although these transitions are typically dominant, the others are certainly non-negligible.

\citet{valdes08} present detailed results for the high-energy limit and offer alternate fitting functions to those from \citet{shull85}; Figure~\ref{fig:high-energy} also includes a comparison to their results (dotted curves).  The agreement with ours is not nearly as good as for \citet{shull85}.  They find a much lower rate of heating at moderate and large ionized fractions (by up to $\sim 20\%$ at $x_i \sim 0.01$) and correspondingly larger rates of excitation and (especially) ionization.  The agreement is particularly poor at $x_i \la 1$ (although the practical importance of this regime is fairly small).  The source of this discrepancy is unclear; our excitation cross-sections match closely in the high-energy limit \citep{stone02}, and our ionization cross-sections are in good agreement over the entire energy range (e.g., comparing the CCC database to \citealt{kim94}).

Another difference between our results and \citet{valdes08} is in the fraction of excitation energy deposited as \lya photons.  As shown in Figure~\ref{fig:excite-lya}, we find $f_{\rm Ly\alpha}/f_{\rm excite} \approx 0.8$ throughout the entire energy range.  In contrast, \citet{valdes08} find $f_{\rm Ly\alpha}/f_{\rm excite} \approx 0.7$.  This discrepancy probably results from their inclusion only of excitations to the $np$ sublevels; the mixing fractions for the (rarer) excitations to the other sublevels can be larger, especially at $n=3$ (where atoms in the 3$p$ state cannot decay through \lyans, but those in the $3s$ or $3d$ states must).  

\begin{figure*}
\begin{center}
\resizebox{8cm}{!}{\includegraphics{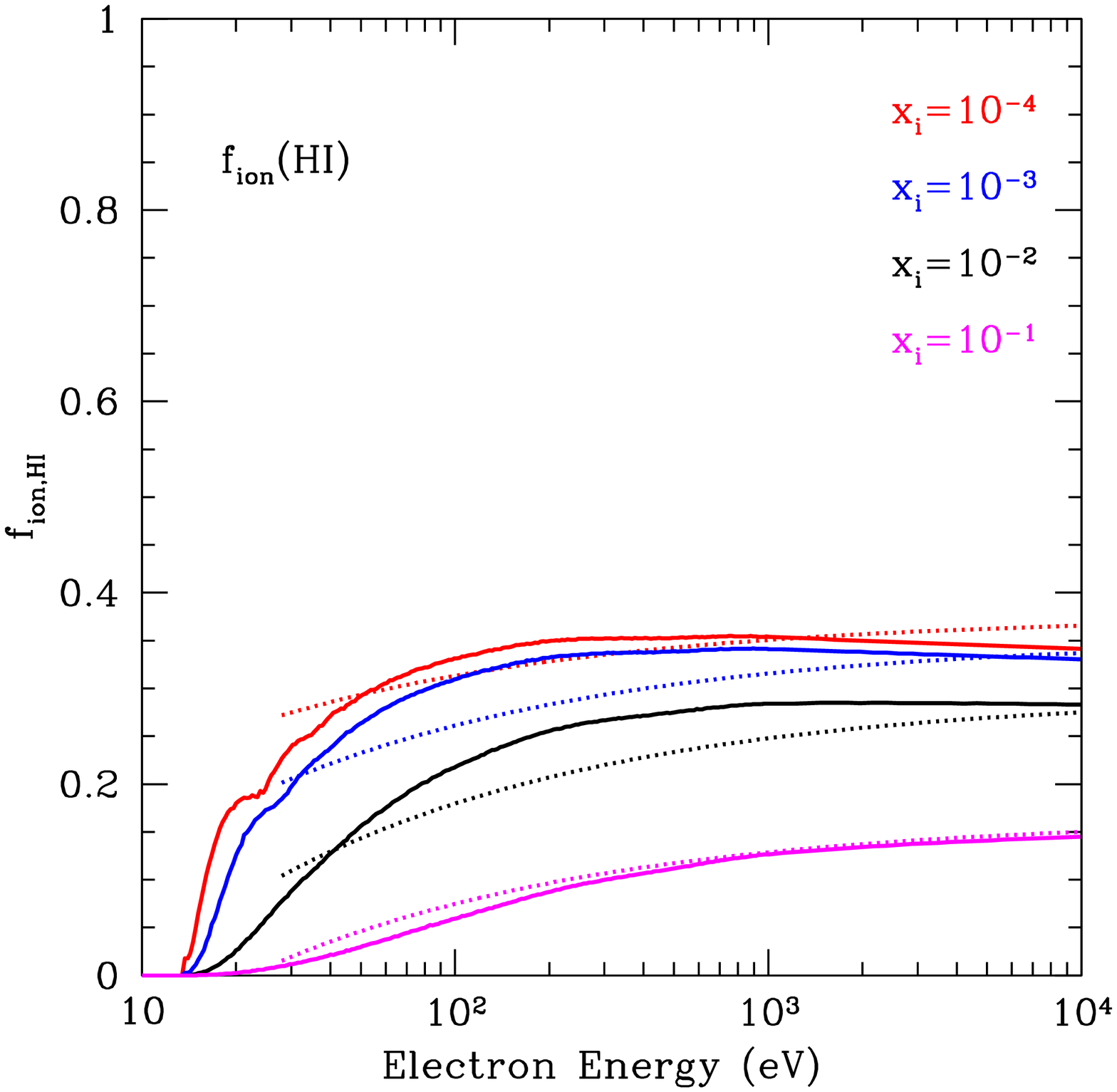}}
\hspace{0.13cm}
\resizebox{8cm}{!}{\includegraphics{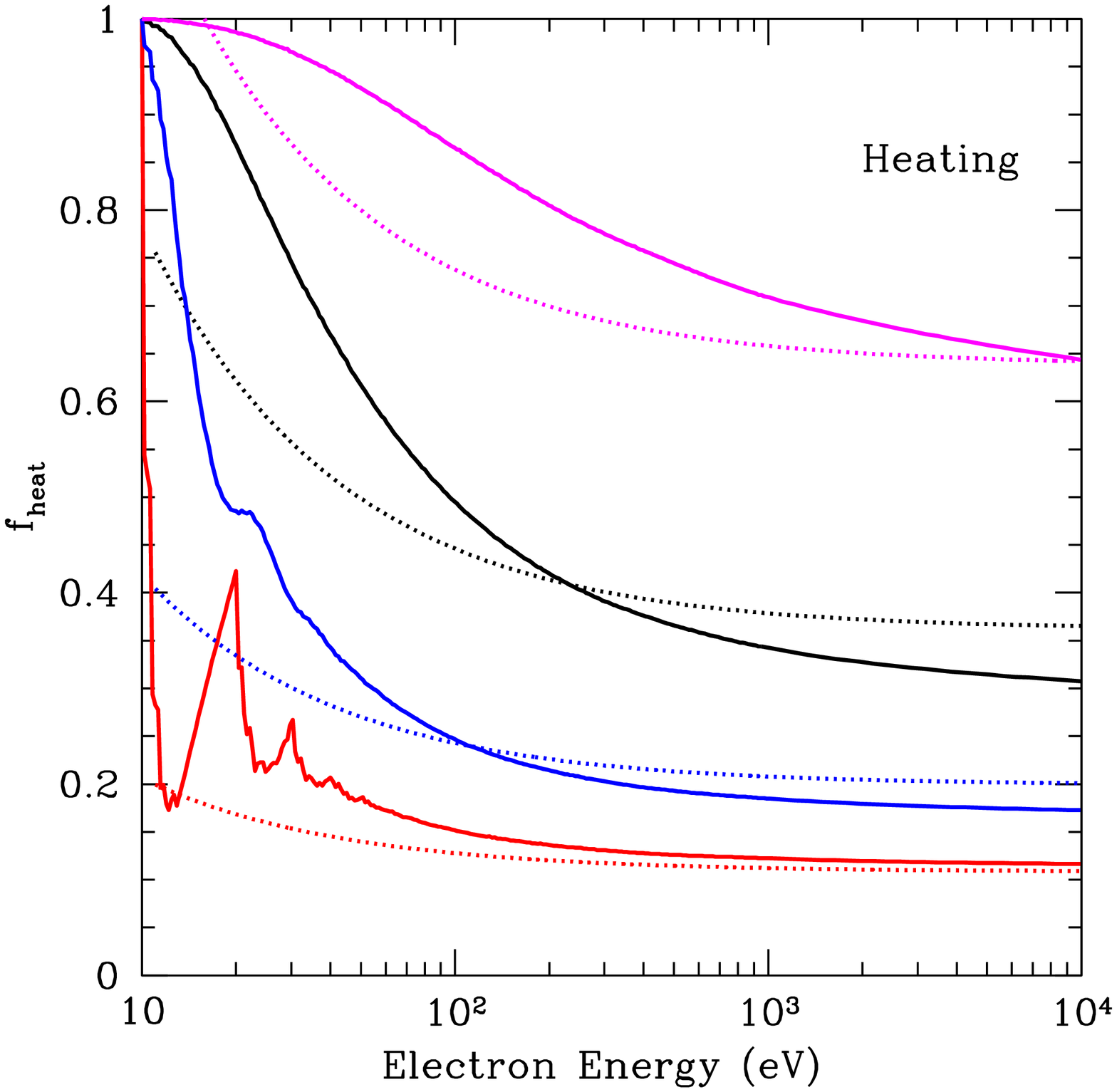}}
\end{center}
\caption{Energy deposition fractions in HI ionization (\emph{left}) and heating (\emph{right}) as a function of electron energy.  The curves in each panel show results for several different IGM ionized fractions, with $\log x_i = -4,\,-3,\,-2,$ and $-1$, from top to bottom.  The solid curves show our calculations; the dotted curves show the fits from \citet{ricotti02}; see eqs.~(\ref{eq:gnedin-fit1}-\ref{eq:gnedin-fit2}) in the text. }
\label{fig:fit-comp}
\end{figure*}

Over lower energies, the agreement with previous results is less clear. \citet{valdes08} examined only the high-energy limit, but \citet{shull85} studied the entire range. They presented few numerical results at lower energies, but \citet{ricotti02} provided the following fits to their results (see also \citealt{volonteri09}):
\bqa
f_{\rm ion,HI}^R & \approx & - 0.69 \left( {28 \eV \over E} \right)^{0.4} x_i^{0.2} (1 - x_i^{0.38})^2 \nonumber \\
& & + 0.39 (1-x_i^{0.41})^{1.76}  \qquad E>28 \eV \label{eq:gnedin-fit1}
\\
f_{\rm heat}^R & \approx & 3.9811 \left( {11 \eV \over E} \right)^{0.7} x_i^{0.4} (1 - x_i^{0.34})^2 \nonumber \\
& & + [1 - (1.0-x_i^{0.27})^{1.32} ] \qquad E > 11 \eV,
\label{eq:gnedin-fit2}
\eqa
and zero otherwise.  The terms without energy dependence describe the high-energy limiting values discussed above; the final term in each expression was added to account for the energy dependence, at least roughly.

Figure~\ref{fig:fit-comp} compares these expressions to our detailed results over the entire relevant energy range.  The left panel shows $f_{\rm ion,HI}$, while the right panel shows $f_{\rm heat}$.  The sets of curves take $\log x_i = -4,\,-3,\,-2,$ and $-1$, from top to bottom, with the solid curves showing our results and the dotted curves using the fits.  Although (for the most part) the fits are reasonably accurate at $E>1 \keV$,\footnote{The largest deviation in the high energy limit, of $\sim 6\%$ in absolute terms, occurs in $f_{\rm heat}$ when $x_i=0.01$.  Note, however, that our result actually agrees with the exact result from \citet{shull85} well (0.32 for both); the error comes from the fitting function itself.}  they provide a relatively poor match at lower energies.  We therefore caution against these fits for high-accuracy work. 

The advantage of equations~(\ref{eq:gnedin-fit1}-\ref{eq:gnedin-fit2}) is that the energy and $x_i$ dependence are separable, which simplifies their application to numerical work.  Unfortunately, Figures~\ref{fig:ion-heat} and~\ref{fig:excite-lya} show that the functional dependence over energy can change quite substantially with $x_i$ (even leaving aside the features from the ionization and excitation thresholds).  We suspect this is why a single, separable fitting function fails, and we have not attempted to find another form.  Instead, we recommend interpolating the exact results.

\section{An example application: discrete sources in the early universe}
\label{example}

\begin{figure*}
\begin{center}
\resizebox{8cm}{!}{\includegraphics{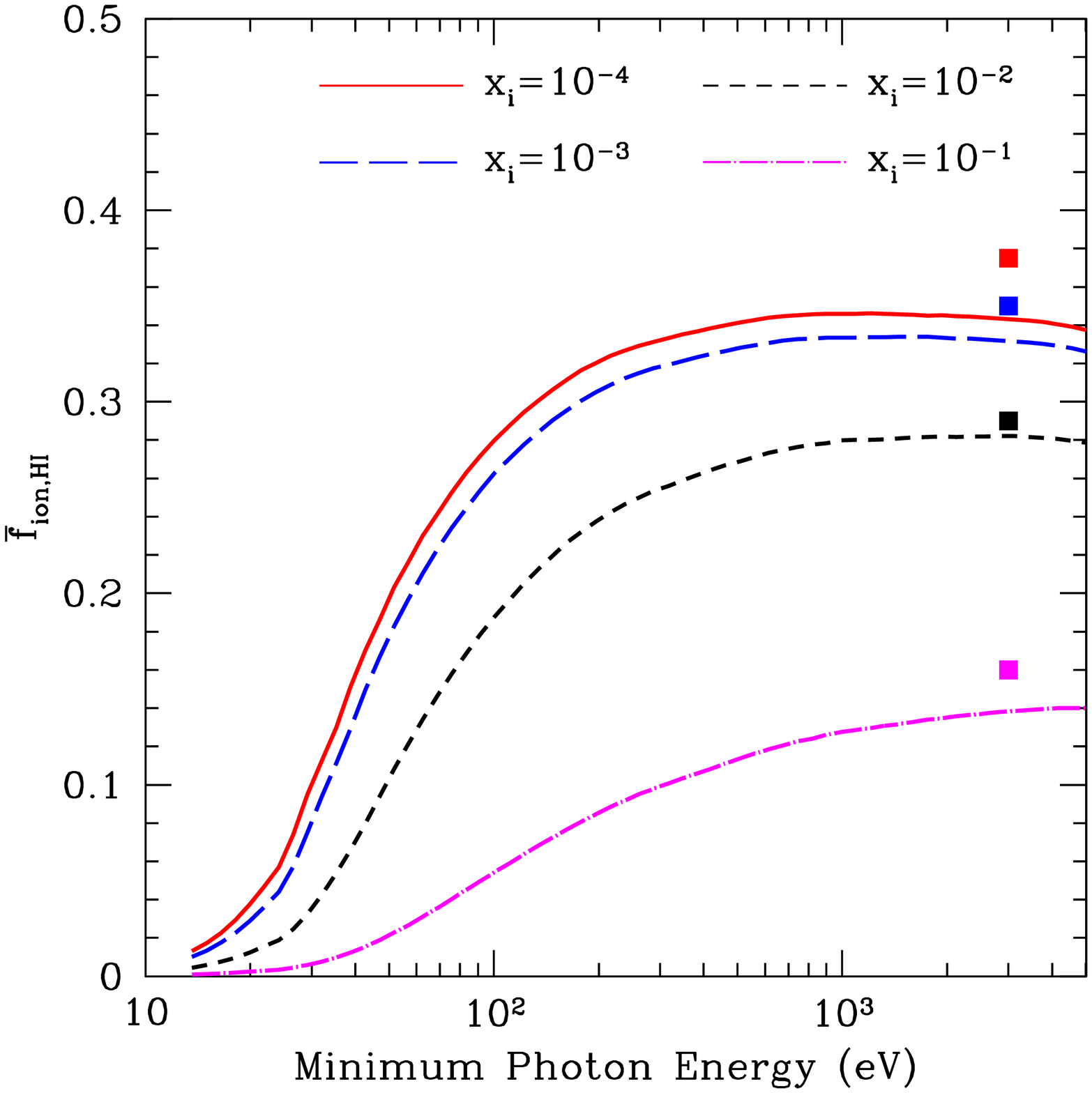}}
\hspace{0.13cm}
\resizebox{8cm}{!}{\includegraphics{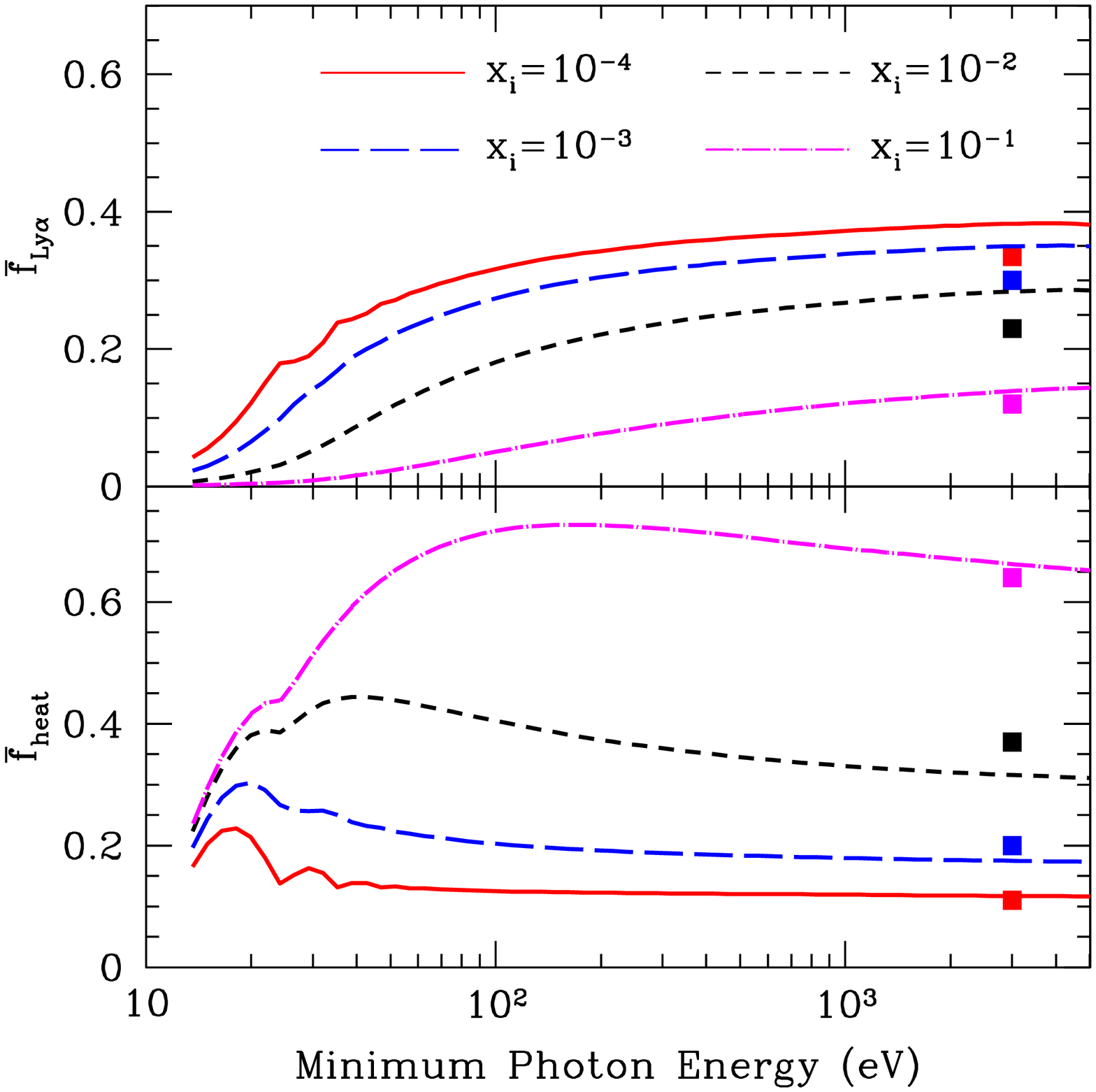}}
\end{center}
\caption{Energy deposition fractions for the secondary electrons integrated over sample spectra (with $\alpha=1.5$), assuming optically thin absorption.  The left panel shows $\bar{f}_{\rm ion,HI}$, the upper right panel shows $\bar{f}_{\rm Ly\alpha}$, and the lower right panel shows $\bar{f}_{\rm heat}$.  The solid, long-dashed, short-dashed, and dot-dashed curves take $\log x_i=-4,\,-3,\,-2,$ and $-1$, respectively.  The solid squares show the high-energy fit from \citet{shull85}, with the same ordering from top to bottom.}
\label{fig:int-spec}
\end{figure*}

As a simple example of the utility of our results, we briefly describe here an application to X-ray heating by discrete sources in the high-$z$ universe.  Suppose a source emitting ultraviolet and X-ray photons with a power law spectrum, $L_\nu \propto \nu^{-\alpha}$, is embedded in the IGM.  We will take $\alpha=1.5$ for concreteness.  The photons will gradually be absorbed as they stream through the neutral gas surrounding the source: ultraviolet photons just above the ionization edge have a very short mean free path, but higher-energy photons can reach much larger distances.  The comoving mean free path of an X-ray photon with energy $E_\gamma$ is 
\bq
\lambda_X \approx  4.9 x_i^{-1/3} \left( { 1 + z \over 15} \right)^{-2} \left( {E_\gamma \over 300 \eV} \right)^3 \Mpc.
\eq

Thus, for a point at a finite distance from our source, all photons with energies $E_\gamma < E_{\rm min}$ will be strongly attenuated, while those above this threshold will be more or less unaffected.  We approximate this situation by assuming zero transmission below $E_{\rm min}$ and complete transmission above that threshold.  We then compute the net fraction of absorbed energy that is deposited as heat:\footnote{By integrating over $\sigma$ here, we are implicitly assuming the optically thin limit for the surviving photons.  If a system is optically thick, it will weight high-energy photons more strongly.}
\bq
\bar{f}_{\rm heat} = \frac{ \sum_i \int_{E_{\rm min}}^\infty dE_\gamma \, E_\gamma^{-\alpha-1} n_i \sigma_i(E_\gamma) (E_\gamma - E_{i})  f_{\rm heat}}
{\sum_i \int_{E_{\rm min}}^\infty dE_\gamma \, E_\gamma^{-\alpha} n_i \sigma_i(E_\gamma)},
\eq
where the sum is over all species $i$ and $\sigma_i$ is the photoionization cross-section \citep{verner96}.  Here the numerator is the integral (over all photon energies $E_\gamma$) of the ionization rate multiplied by the fraction of the secondary electron's energy $(E_\gamma - E_i)$ deposited as heat.  The denominator normalizes $\bar{f}$ to the total energy deposition rate.  We can compute $\bar{f}_{\rm ion,HI}$ and $\bar{f}_{\rm Ly\alpha}$ in a similar fashion, replacing $f_{\rm heat}$ with $f_{\rm ion,HI}$ or $f_{\rm Ly\alpha}$.  (Note that $\bar{f}$ still does \emph{not} include the energy deposited by the initial ionization.  This is why the sum of these fractions is much smaller than unity at low energies.)

The curves in Figure~\ref{fig:int-spec} show these two quantities at several different ionized fractions ($\log x_i=-4,\,-3,\,-2,$ and $-1$, from top to bottom in the left hand panel).  The solid squares show the asymptotic high-energy estimates from the \citet{shull85} fitting formulae.  (These are the terms in eqs.~\ref{eq:gnedin-fit1}-\ref{eq:gnedin-fit2} without any energy dependence.)

These fitting formulae provide reasonable order-of-magnitude estimates so long as $E_{\rm min} \ga 300 \eV$ (or slightly higher if $x_i \sim 0.1$), although they systematically underestimate $\bar{f}_{\rm Ly\alpha}$ and $\bar{f}_{\rm ion,HI}$.  These deviations are most significant at small ionized fractions, where heating is least important.  We therefore again recommend interpolation of the exact results for high-accuracy work, especially if the heating by soft X-rays $E \la 300 \eV$ is included.

\section{Discussion}
\label{disc}

Using a Monte Carlo model, we have re-examined the fate of fast electrons scattering through a background gas of primordial origin.  We included electron-electron scattering as well as collisional ionization and excitation of HI, HeI, and HeII, explicitly tracking all levels up to $n=4$ and using an analytic extrapolation to higher levels.  We separately followed all excitations producing HI \lya photons, which can be important in modeling the observable properties of the IGM at high redshifts (see \citealt{kuhlen06-21cm} and \citealt{furl06-review}, for example) and have not been explicitly tracked previously except at the highest energies.  We used recent calculations of ionization and excitation cross-sections at $E<1 \keV$ and extrapolated to higher energies using the Bethe approximation.

In highly neutral gas ($x_i \la 10^{-3}$), we found that $\sim 20\%$ of the electron energy is deposited as heat, with the remainder split roughly equally between ionization and excitation.  In this regime, the results are not strongly sensitive to $x_i$, at least at high energies, because most of the heating comes from secondary electrons, with energies below $10 \eV$.  At higher $x_i$, the heating fraction rises rapidly, exceeding $\sim 65\%$ by $x_i \sim 0.1$.  We find that the excitation and ionization energy deposition rates are always comparable, and that $\sim 80\%$ of the excitation energy goes into HI \lya regardless of electron energy and $x_i$.

Although our calculations used parameters appropriate to the low-density IGM, our results may also be applied to denser systems, because the density only enters through the Coulomb logarithm affecting the electron-electron scattering rate.  Varying the density of target atoms by many orders of magnitude only affects the energy deposition fractions by a few percent in absolute terms.  We also note that, when collective plasma effects are included, the background temperature becomes irrelevant for the electron energies under consideration \citep{schunk71,xu91}.

For the most part, our results agree with previous estimates from \citet{shull85} and \citet{xu91}, although we have found some discrepancies with the commonly-used fitting formulae from the former.  In general, at high energies their results slightly underestimate the importance of collisional excitation but are otherwise accurate.  At lower energies, the differences in cross-sections become more important and the discrepancies increase (at least with reference to the fitting formulae of \citealt{ricotti02}).  However, our results show substantial differences with those of \citet{valdes08}, who examined the high-energy limit with a Monte Carlo model similar to ours.  These discrepancies are especially large at moderate and high ionized fractions, where we find substantially more heating and less ionization and excitation.  We also find that a higher fraction of excitation energy is deposited in HI \lya photons.  The latter is probably due to our different treatments of collisional excitation (in particular, their neglect of excitations to states other than the $np$ sublevels), but the source of the former is unclear.

In any case, we advocate interpolation of the exact results when high accuracy is necessary, especially because the energy dependence is quite significant.  Our detailed numerical results are available upon request, including tables for the energy deposition fractions at a variety of ionized fractions and C code to interpolate to arbitrary ionized fractions and electron energies.

\vskip 0.5in

We thank M. Valdes for helpful comments. SRF was partially supported by the NSF through grant AST-0829737, by the David and Lucile Packard Foundation, and by NASA through the LUNAR program. The LUNAR consortium (http://lunar.colorado.edu), headquartered at the University of Colorado, is funded by the NASA Lunar Science Institute (via Cooperative Agreement NNA09DB30A) to investigate concepts for astrophysical observatories on the Moon.  SJS was supported by a Research Experiences for Undergraduates grant at UCLA, NSF PHY-0552500.  We thank Agner Fog for making a public version of the Mersenne Twister algorithm available and Igor Bray and Yuri Ralchenko for making the CCC database available electronically.

\bibliographystyle{mn2e}
\bibliography{Ref_composite}

\end{document}